\documentclass{PoS}

\usepackage{graphicx}
\usepackage{epstopdf}
\usepackage{amssymb}
\usepackage{amsmath}
\usepackage{booktabs}
\usepackage{float}

\usepackage{adjustbox}

\usepackage{epstopdf}

\newcommand{\GeV}{~\mathrm{GeV}}
\newcommand{\TeV}{~\mathrm{TeV}}

\newcommand{\tinymath}[1]{{{\mbox{\tiny $#1$}}}}
\newcommand{\zp}{Z^{\prime}}
\newcommand{\Mzp}{M_{Z^\prime}}
\newcommand{\mzp}{M_{\tilde{Z}^\prime}}
\newcommand{\uonep}{\ensuremath{U(1)^\prime}}
\newcommand{\gz}{\ensuremath{g_{\tinymath{Z^\prime}}}}
\newcommand{\be}{\begin{equation}}
\newcommand{\ee}{\end{equation}}

\title{Imposing LHC constraints on the combined Anomaly and $\zp$ Mediation Mechanism of Supersymmetry Breaking}

\ShortTitle{Combined Anomaly and $\zp$ Mediation of SUSY breaking}

\author{\speaker{Joydeep Roy}\\
        Wayne State University\\
        Detroit, U.S.A\\
        E-mail: \email{joydeep.roy@wayne.edu}}

\abstract{Combining anomaly with $Z^\prime$ mediation allows us to solve the tachyonic problem of the former and avoid fine tuning in the latter. This model includes an extra $U(1)^\prime$ gauge symmetry and extra singlet scalar $S$ which provides a solution to the `$\mu$ problem' of the Minimal Supersymmetric Standard Model (MSSM). The low-energy particle spectrum is calculated from the UV inputs using the Renormalization Group Equations. The benchmark points considered in the original model, suggested before the Higgs discovery, predicted a Higgs mass heavier than the generic MSSM value.  In 2012, the Higgs  particle was discovered and found to have a mass of 125 GeV. Therefore, we can use that value and other current LHC data to scan the parameter space and update the predictions of the model, in particular the mass of the $Z^\prime$ gauge boson.}

\FullConference{38th International Conference on High Energy Physics\\
		3-10 August 2016\\
		Chicago, USA}

\begin{document}

\section{Introduction}
One of the prime motivation to conceive the idea of supersymmetry (SUSY) was to stabilize the Higgs mass and solve the hierarchy problem. Several decades of study and development of this theory has revealed that supersymmetry must be a broken symmetry. It is believed that SUSY is broken at very high energy level, known as `hidden sector' and then it is `communicated' to the Electroweak (EW) scale, known as `visible sector'. Therefore the most important questions in supersymmetric theory are, `how the sypersymmetry is broken and how this breakdown is communicated between two sectors' ?

\section{SUSY-breaking mechanisms}
There are several SUSY-breaking mechanisms available in the literature. These are known as Gauge-mediated supersymmetry breaking (GMSB) \cite{Dine:1981gu Nappi:1982hm Dine:1993yw Dine:1994vc}, Planck-scale-mediated supersymmetry breaking (PMSB) \cite{Chamseddine:1982jx Barbieri:1982eh Hall:1983iz Ibanez:1982ee Ohta:1982wn}, Extra-dimensional mediated supersymmetry breaking (``XMSB") \cite{Kaplan:1999ac Chacko:1999mi} or Anomaly-mediated supersymmetry breaking (AMSB) \cite{Randall:1998uk Giudice:1998xp}. The extension of the Minimal Supersymmetric Standard Model (MSSM) via a $\uonep$ gauge group can also be considered as a mediator of SUSY breaking where the $\uonep$ vector multiplet communicate between two sectors \cite{Langacker:2007ac}. 

\subsection{$\zp$ mediation mechanism of SUSY breaking}

$\zp$ mediation of SUSY breaking is a mediation mechanism in which both the hidden and the visible sectors are charged under a new $\uonep$ gauge interaction. The associated boson of this $\uonep$ extension is the $\zp$ gauge boson which is produced when $\uonep$ gauge group is broken at the TeV scale \cite{Langacker:2008yv}. Since a $\uonep$ can couple to both MSSM sector and the hidden sector, the $\zp$ has been considered as a mediator of the Supersymmetry (SUSY) breaking mechanism in several theoretical models such as \cite{Langacker:2007ac,Langacker:2008ip}. Though this scenario is often referred as $\zp$ mediation, it can be thought of as a $\zp$ gaugino mediated mechanism. 

\subsubsection{General features of $\zp$ mediation}
The general features of the original $\zp$ mediation mechanism \cite{Langacker:2007ac,Langacker:2008ip} can be summarized as follows:
\begin{itemize}
\item A new $\uonep$ gauge symmetry is introduced under which all fields are charged. These charges are family universal.
\item This $\uonep$ gauge group couples to both the visible and hidden sectors.
\item It a possible solution of ``$\mu$-problem" by introducing the SM singlet superfield $S$ which is charged under $\uonep$ so that the superpotential term $SH_uH_d$ is allowed.
\item To cancel the new anomalies the following ``exotic" matter are introduced:
\begin{itemize}
\item 3 pairs of colored, $SU(2)_L$ singlet exotics $D,D^c$ with
  hypercharge $Y_D=-1/3$ and $Y_{D^c}=1/3$.
\item 2 pairs of uncolored $SU(2)_L$ singlet exotics $E,E^c$ with
  hypercharge $Y_E=-1$ and $Y_{E^c}=1$.
\end{itemize}
\item The exotic fields can couple to $S$, namely the superpotential terms $S D D^c$ and $S E E^c$ are allowed.
\end{itemize}

Finally the superpotential is given by
\begin{eqnarray}\label{Superpotential}
W&=& y_u  {H}_u {Q} {u}^c + y_d  {H}_d {Q} {d}^c + y_e  {H}_d {L} {e}^c+y_\nu  {H}_u {L} {\nu}^c
\\\nonumber
&+&\lambda  {S}  {H}_u  {H}_d+ y_D\, S\left( \sum_{i=1}^{3}  D_i D_i^c \right) +y_E\, S\left (\sum_{j=1}^{2} E_j E_j^c\right).
\end{eqnarray}

\subsubsection{Features of mass spectrum}

Since it is assumed that all the chiral superfields in the visible sector are charged under $\uonep$, all the corresponding scalars receive mass terms at 1-loop of order 
\be \label{eqn:Scalar mass}
m^2_{\tilde{f_i}} \sim \frac{\gz^2 Q^2_{f_i}}{16 \pi^2}  \,\, M^2_{\tilde{\zp}} \,\, \mathrm{log}  \bigg(\frac{\Lambda_S}{M_{\tilde{\zp}}}\bigg),
\ee 
where $\gz$ is the $\uonep$ gauge coupling and $Q_{f_i}$ is the $\uonep$ charge of fermion $f_i$.

The $SU(3)_C \times SU(2)_L \times U(1)_Y$ gaugino masses can be generated at 2-loops since they do not directly couple to the $\uonep$ \cite{Langacker:2008ip},
\begin{eqnarray}
\label{eqn:gauginomass}
&\quad& \nonumber \\
{M}_a~~ &\sim& \parbox[t]{5cm}{\vspace{-1cm}
\includegraphics[scale=0.45]{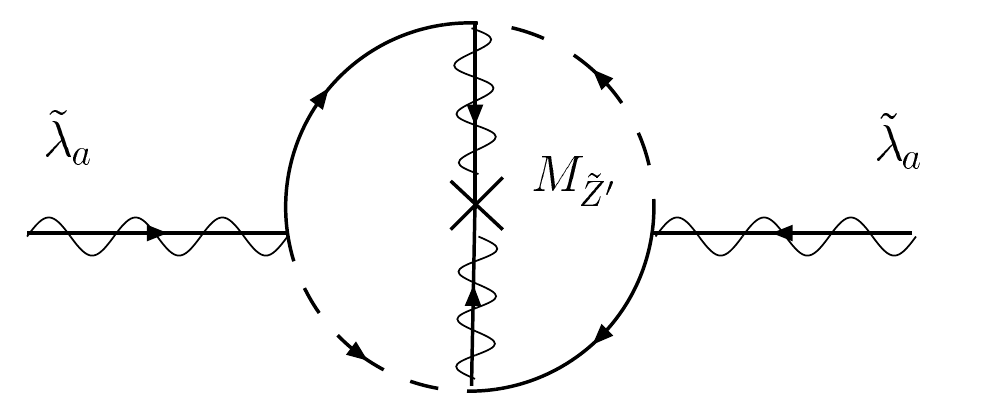} }  \nonumber \\
&\sim& ~~\frac{\gz^2 g_a^2}{(16\pi^2)^2} \mzp
\log\left(\frac{\Lambda_S}{\mzp} \right),
\end{eqnarray}
where $g_a$ is the gauge coupling for the gaugino $\tilde{\lambda}_a$, and the
internal line is the sum over the
chiral supermultiplets charged under the $a^{th}$ gauge group. 

From equations \eqref{eqn:Scalar mass} and \eqref{eqn:gauginomass} we see that, for ${M}_a\gtrsim 100\GeV$ (LEP direct searches bound) and $\gz$ is of electroweak strength,
\be
{m}_{\tilde{f}_i} \sim \frac{(4\pi)^3}{\gz g_a^2} M_a \sim 100\TeV.
\ee

Therefore we are left with two possibilities. First, we can choose Gauginos to be at EW scale ($\sim 100-1000 \GeV$) which implies heavy scalars ($\sim 100 \TeV$) and very heavy $\zp$-gaugino mass ($\mzp \sim 1000\TeV$) and to obtain EW symmetry breaking at the observed scale fine tuning is needed. Other possibility is to choose Scalars at EW scale ($\sim 100-1000 \GeV$). In this scenario the gauginos are too light and must acquire mass from other mechanism. The obvious candidate is gravity mediation which gives a contribution to the gaugino mass of order $F/M_P$, where $F$ is the SUSY breaking scale and $M_P$ is the Planck scale. Choosing higher values of $F$ can have significant contribution to the gaugino mass. Similarly, it was shown in \cite{deBlas:2009vx} that anomaly mediation (AMSB) can also contribute significantly to the gaugino mass. We will follow this second possibility according to \cite{deBlas:2009vx}.

\subsection{Anomaly mediation mechanism of SUSY breaking }
In anomaly mediation the gauge supermultiplet fields are assumed to be confined to the MSSM brane and the SUSY breaking effect is communicated due to the supergravity (SUGRA) effect. But at tree-level this SUGRA effect doesn't give rise to soft SUSY breaking in the observable sector. The masses of soft SUSY breaking parameters are generated at loop level by the anomalous violation of local superconformal invariance \cite{Randall:1998uk Giudice:1998xp}.

\subsubsection{Features of mass spectrum}
In anomaly mediation the masses are proportional to the gravitino mass ($m_{3/2}$) as this is the mediator between two sectors. The soft scalars get masses in 2-loops \cite{Randall:1998uk Giudice:1998xp},
\be \label{eqn:Scalar mass in AMSB}
m_S^2 = -\frac{1}{4}\bigg(\frac{\partial \gamma}{\partial g} \beta_g +  \frac{\partial \gamma}{\partial y} \beta_y \bigg) m_{3/2}^2
\ee
where $\gamma = d\mathrm{ln} Z_Q / d\mathrm{ln} \mu$ ,$ \beta_g = d\mathrm{g} /  d\mathrm{ln} \mu$, $ \beta_y = d\mathrm{y} /  d\mathrm{ln} \mu$, $g$ and $y$ are the gauge couplings  and $Z$ is some function of high energy scales.

The gauginos in this mechanism get mass in 1-loop,
\be \label{eqn:Gaugino mass in AMSB}
M_a = \frac{\beta_g}{g}  m_{3/2}.
\ee
One of the features of the mass spectrum as well as drawbacks of this AMSB is the presence of `negative' slepton masses due to small Yukawa couplings.

\section{Combining Anomaly and $\zp$ mediation mechanism}
$\zp$ gaugino and anomaly mediation are similar in the sense that both are flavor diagonal. Also, comparing the soft mass spectrum of both, as discussed above, it is clear that the scale of the soft parameters is set by one dimensionful parameter for each mechanism. For $\zp$-gaugino mediation this parameter is the $\zp$-gaugino mass $\mzp$, for the anomaly mediation it is the gravitino mass $m_{3/2}$ and they are related by
\be 
\frac{m_{3/2}}{\mzp} \sim 4\pi.
\ee
Such a mild hierarchy between the two mediators can be realized and therefore both, $\zp$-gaugino and anomaly can be combined to avoid the fine tuning problem for the former and address the negative `slepton' mass problem of the latter as shown in \cite{deBlas:2009vx}.

\section{Specific illustration point}

To get a realistic spectrum we need to have some dimensionful as well as some dimensionless input parameters. Such parameters were chosen for two specific illustration points in \cite{deBlas:2009vx}. The dimensionless parameters for one of the points are following:
\begin{eqnarray}
\uonep \mbox{ gauge coupling (at } \Lambda_{S}) \quad \mbox{and charges} &:&\quad \gz=0.45 \quad \mbox{and} \quad Q_{H_u} = -\frac{2}{5},\quad Q_Q = -\frac{1}{3}\\
\mbox{Superpotential couplings (at } \Lambda_{\rm EW})
&:& y_t = 1,y_b = 0.5,y_{\tau} = 0.294,\lambda=0.1,\, y_D=0.3,\, y_E=0.5\nonumber.
\end{eqnarray}

The relevant scalar and gaugino mass spectrum for this illustration point are shown in Tables
\eqref{table:Higgs Masses} and \eqref{table:Gaugino masses}. The stop masses are found to be $m_{\tilde{t_{1}}} = 0.695 \TeV$  and $m_{\tilde{t_{2}}} = 3.16 \TeV$. The $\zp$ gauge boson mass is found to be $\Mzp = 2.78 \TeV$. The scalar masses are calculated including the radiative corrections. The other illustration point similarly yields a $\zp$ gauge boson mass, $\Mzp = 5.68$ TeV and Higgs mass $m_{h^0} = 0.142$ TeV.

\begin{table}[H]
\begin{minipage}{2.0 in}
\begin{tabular}[t]{ | c | c | c |}
\hline
$m_{h^0}$ & $m_{H_1^0}$ & $m_{H_2^0}$ \\
\cline{1-3}
0.138 TeV  & 2.79 TeV & 4.78 TeV \\
\hline
\end{tabular}
\caption{Higgs masses }\label{table:Higgs Masses}
\end{minipage}\qquad
\hspace{2.2 cm}
\begin{minipage}{2.0 in}
\begin{tabular}{ | c | c | c |}
\hline
Wino & Gluino & Bino \\
\cline{1-3}
0.279 TeV  & 0.399 TeV & 1.17 TeV \\
\hline
\end{tabular}
\caption{Gaugino masses}\label{table:Gaugino masses}
\end{minipage}
\end{table}

\section{Present work}
Currently, in order to put constraints on the mass spectrum we are looking for the Leading Order (LO) cross-section at LHC relevant for Drell-Yan process of $\zp$ production and decay \cite{Paz & Roy}. This cross-section can be parameterize in the following way \cite{Carena:2004xs}
\be\label{eqn:Cross-section}
\sigma_{l^+l^-}^{LO} = \frac{\pi}{48 s}\big[c_u w_u(s,\Mzp) + c_d w_d(s,\Mzp)\big],
\ee
where 
\begin{eqnarray}
c_{u,d} = \frac{\gz^2}{2}\big[\big(g_V^{u,d}\big)^2 + \big(g_A^{u,d}\big)^2\big] \quad \mbox{and}\quad w_{u,d}(s,\Mzp) = \int_0^1dx_1 f_{u,d}(x_1)\int_0^1dx_2 f_{\bar{u},\bar{d}}(x_2)\delta(\frac{\Mzp^2}{s} - x_1x_2).\nonumber
\end{eqnarray}
The vector and axial couplings $(g_{V,A}^f)$ are related to the chiral couplings by the following relation, $g_{V,A}^f = \epsilon_L^f \pm \epsilon_R^f $. We see from \eqref{eqn:Cross-section}, that all the model dependence of cross-section is contained in $c_u$ and $c_d$. Therefore for any benchmark model, the collider limits on $\zp$ mass can be obtained by contours in $c_u - c_d$ plane as shown in the following plot.

\begin{figure}[H]
\centering
{\includegraphics[width=3.6in,height=2.5in, angle=0]{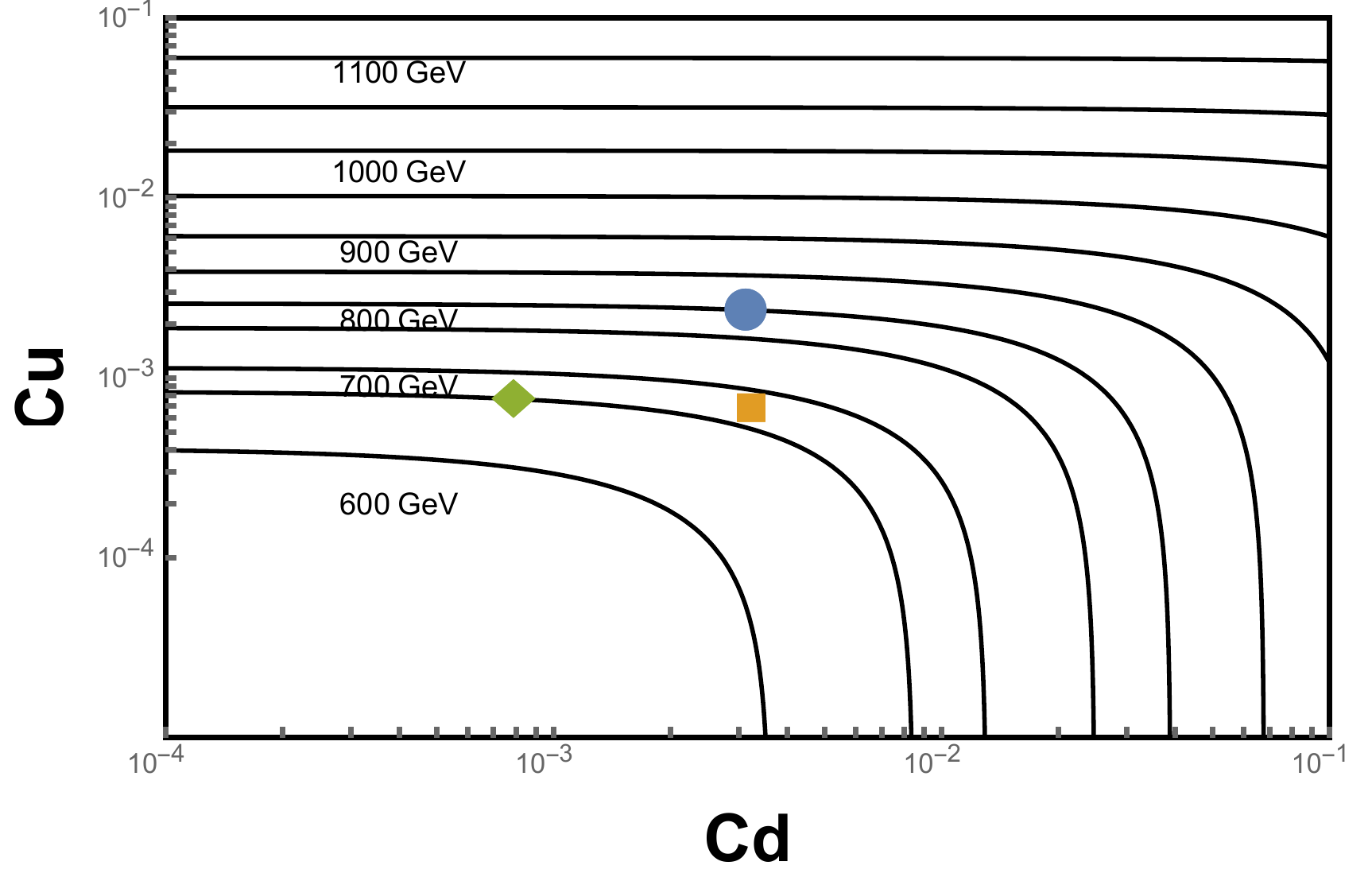}}
\caption{\label{fig:Plot for zp mass limits} {The $\zp$ mass limits obtained for the benchmark models using D0 collaboration data \cite{Abazov:2010ti}. The 'Blue dot', 'Orange box' and 'Red diamond' represent $\uonep_{SSM}$, $\uonep_{\chi}$ and $\uonep_{\psi}$ models respectively  
}}
\end{figure}

In Figure \eqref{fig:Plot for zp mass limits} we show the present $\zp$ mass limits obtained for different benchmark models, using D0 collaboration data \cite{Abazov:2010ti} for the cross-sections and \eqref{eqn:Cross-section} .

\section{Future work and outlook}
From the current LHC data we have imposed the constraints on the $\zp$-boson mass. Since the $\zp$ mass, $\Mzp \approx \gz Q_S \langle S \rangle $ and $c_{u,d} \varpropto \gz^2$, it is extremely important to choose suitable $\gz$ and $\langle S \rangle $ to be in the
experimentally allowed region.

We also plan to use the observed Higgs mass ($125 \GeV$) as an input to scan the parameter space, update the whole mass spectrum and put constraints on the gluinos and stops masses \cite{Paz & Roy}.


\begin{thebibliography}{99}

\bibitem{Dine:1981gu Nappi:1982hm Dine:1993yw Dine:1994vc} 
  M.~Dine and W.~Fischler,
  Phys.\ Lett.\  {\bf 110B}, 227 (1982).
  C.~R.~Nappi and B.~A.~Ovrut,
  M.~Dine and A.~E.~Nelson,
  Phys.\ Rev.\ D {\bf 48}, 1277 (1993). 
  M.~Dine, A.~E.~Nelson and Y.~Shirman,
  Phys.\ Rev.\ D {\bf 51}, 1362 (1995)
  
  
\bibitem{Chamseddine:1982jx Barbieri:1982eh Hall:1983iz Ibanez:1982ee Ohta:1982wn} 
  A.~H.~Chamseddine, R.~L.~Arnowitt and P.~Nath,
  Phys.\ Rev.\ Lett.\  {\bf 49}, 970 (1982).
  R.~Barbieri, S.~Ferrara and C.~A.~Savoy,
  Phys.\ Lett.\  {\bf 119B}, 343 (1982).
  L.~J.~Hall, J.~D.~Lykken and S.~Weinberg,
  Phys.\ Rev.\ D {\bf 27}, 2359 (1983). 
  L.~E.~Ibanez,
  Phys.\ Lett.\  {\bf 118B}, 73 (1982).
  N.~Ohta,
  Prog.\ Theor.\ Phys.\  {\bf 70}, 542 (1983).
  
\bibitem{Kaplan:1999ac Chacko:1999mi} 
  D.~E.~Kaplan, G.~D.~Kribs and M.~Schmaltz,
  Phys.\ Rev.\ D {\bf 62}, 035010 (2000) 
  Z.~Chacko, M.~A.~Luty, A.~E.~Nelson and E.~Ponton,
  JHEP {\bf 0001}, 003 (2000) 
  
\bibitem{Randall:1998uk Giudice:1998xp} 
  L.~Randall and R.~Sundrum,
  Nucl.\ Phys.\ B {\bf 557}, 79 (1999)
  G.~F.~Giudice, M.~A.~Luty, H.~Murayama and R.~Rattazzi,
  JHEP {\bf 9812}, 027 (1998)
  
\bibitem{Langacker:2007ac} 
  P.~Langacker, G.~Paz, L.~T.~Wang and I.~Yavin,
  Phys.\ Rev.\ Lett.\  {\bf 100}, 041802 (2008)
   

\bibitem{Langacker:2008yv} 
  P.~Langacker,
  Rev.\ Mod.\ Phys.\  {\bf 81}, 1199 (2009)
  
  
\bibitem{Langacker:2008ip} 
  P.~Langacker, G.~Paz, L.~T.~Wang and I.~Yavin,
  Phys.\ Rev.\ D {\bf 77}, 085033 (2008)
  
  
\bibitem{deBlas:2009vx} 
  J.~de Blas, P.~Langacker, G.~Paz and L.~T.~Wang,
  JHEP {\bf 1001}, 037 (2010)
 
  
\bibitem{Paz & Roy} 
  G.~Paz and J.~Roy,
  ``Imposing LHC constraints on the combined Anomaly and $\zp$ Mediation Mechanism of Supersymmetry Breaking''[\emph{in preparation}].
  
\bibitem{Carena:2004xs} 
  M.~Carena, A.~Daleo, B.~A.~Dobrescu and T.~M.~P.~Tait,
  Phys.\ Rev.\ D {\bf 70}, 093009 (2004)
  
\bibitem{Accomando:2010fz} 
  E.~Accomando, A.~Belyaev, L.~Fedeli, S.~F.~King and C.~Shepherd-Themistocleous,
  Phys.\ Rev.\ D {\bf 83}, 075012 (2011)
  
\bibitem{Abazov:2010ti} 
  V.~M.~Abazov {\it et al.} [D0 Collaboration],
  Phys.\ Lett.\ B {\bf 695}, 88 (2011)
  
\bibitem{Khachatryan:2014fba} 
  V.~Khachatryan {\it et al.} [CMS Collaboration],
  JHEP {\bf 1504}, 025 (2015)
  

\end{thebibliography}
\end{document}